\documentclass[journal]{IEEEtran}
\ifCLASSINFOpdf
\else
\fi
\usepackage{amssymb}
\usepackage{amsmath}
\usepackage[pagewise, mathlines, switch]{lineno}
\usepackage{tabularx}
\usepackage{booktabs,rotating}
\usepackage{caption}
\usepackage{subcaption}
\usepackage{threeparttable}
\usepackage{xcolor}
\newcolumntype{Y}{>{\hsize0.5\hsize \arraybackslash}X}

\begin{document}
\bstctlcite{IEEEexample:BSTcontrol}
%
\title{Adversarial Transformer for Repairing Human Airway Segmentation}
%
%
%

\author{Zeyu~Tang, 
        Nan~Yang, 
        Simon Walsh
        and~Guang~Yang,~\IEEEmembership{Senior~Member,~IEEE}
\thanks{This study was supported in part by the BHF (TG/18/5/34111, PG/16/78/32402), the ERC IMI (101005122), the H2020 (952172), the MRC (MC/PC/21013), the Royal Society (IEC/NSFC/211235), the SABER project supported by Boehringer Ingelheim Ltd, the NVIDIA Academic Hardware Grant Program, and the UKRI Future Leaders Fellowship (MR/V023799/1).}
\thanks{S.Walsh and G.Yang are the co-last senior authors. Send correspondence to Y. Nan and G. Yang}
\thanks{Zeyu Tang and Nan Yang are with the National Heart and Lung Institute, Imperial College London, SW7 2BX London, U.K. Zeyu is also with the Department of Bioengineering, Imperial College London, SW7 2AZ London, U.K. (e-mail: zeyu.tang19@imperial.ac.uk; y.nan20@imperial.ac.uk)}
\thanks{Simon Walsh and Guang Yang are with the National Heart and Lung Institute, Imperial College London, SW7 2BX London, U.K., and also with the Royal Brompton Hospital, SW3 6NP London, U.K. (e-mail: s.walsh@imperial.ac.uk, g.yang@imperial.ac.uk).}}

%
%

\markboth{Journal of \LaTeX\ Class Files,~Vol.~14, No.~8, August~2015}%
{Shell \MakeLowercase{\textit{et al.}}: Bare Demo of IEEEtran.cls for IEEE Journals}
%



\maketitle
\begin{abstract}
Discontinuity in the delineation of peripheral bronchioles hinders the potential clinical application of automated airway segmentation models. Moreover, the deployment of such models is limited by the data heterogeneity across different centres, and pathological abnormalities also make achieving accurate robust segmentation in distal small airways difficult. Meanwhile, the diagnosis and prognosis of lung diseases often rely on evaluating structural changes in those anatomical regions. To address this gap, this paper presents a patch-scale adversarial-based refinement network that takes in preliminary segmentation along with original CT images and outputs a refined mask of the airway structure. The method is validated on three different datasets encompassing healthy cases, cases with cystic fibrosis and cases with COVID-19. The results are quantitatively evaluated by seven metrics and achieved more than a 15\% rise in detected length ratio and detected branch ratio, showing promising performance compared to previously proposed models. The visual illustration also proves our refinement guided by a patch-scale discriminator and centreline objective functions is effective in detecting discontinuities and missing bronchioles. Furthermore, the generalizability of our refinement pipeline is tested on three previous models and improves their segmentation completeness significantly.
\end{abstract}


%
\IEEEpeerreviewmaketitle

\section{Introduction}
%
%
%
%
\IEEEPARstart{A}{irway} segmentation from chest computerised tomography (CT) scan serves an important role in the diagnosis and prognosis of pulmonary diseases. Manual segmentation by radiologists is highly time-consuming and error-prone due to the large volume of CT data and complex airway tree structure. To relieve the radiology experts from these tedious manual labelling processes, many automated and semi-automated algorithms are being developed. The performance of traditional segmentation methods, such as thresholding \cite{thresholding} and region-growing \cite{region_growing01}\cite{region_growing02}, have been benchmarked in the EXACT'09 challenge \cite{exact09}, and no algorithm could extract more than an average of 74\% of the total tree length. Recent developments in convolutional neural networks inspired researchers to try models such as U-Net \cite{Unet}, 3-D U-Net \cite{3dUnet}, V-Net \cite{Vnet} and their derivatives \cite{CHARBONNIER201752}\cite{Jin2017}\cite{Meng2017} on the airway segmentation problem. 
\begin{figure}[!hbt]
    \includegraphics[width=0.5\textwidth]{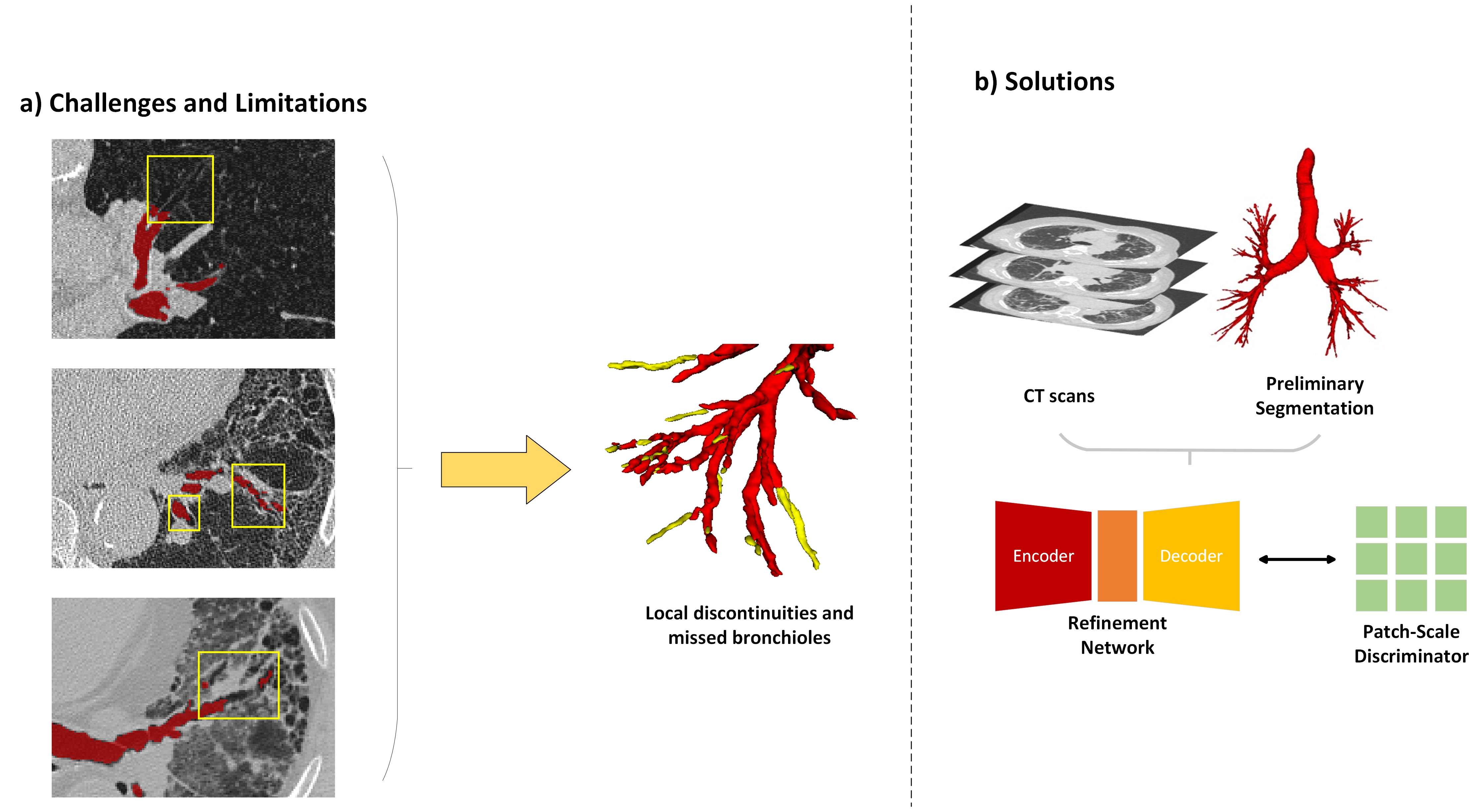}
    \centering
    \caption{Current challenges and solutions of airway segmentation. (a) Examples of blurry boundaries, pathological changes and low resolution (indicated in yellow boxes) leading to discontinued segmentation and missing branches highlighted in yellow. (b) This paper proposes an adversarial-based refinement network to help achieve complete and robust airway segmentation. Patch-scale discriminators are incorporated to penalize local structural errors.}
    \label{fig:challenges_solutions}
\end{figure}
However, despite the improvements in Dice score, CNN-based models still suffer from local discontinuity, especially in the peripheral region of the lung where there is an enormous imbalance between the target distal small airways and the background. Moreover, these deep neural networks are trained on image patches due to the high computation complexity. The small errors in each patch are hard to notice until these patches are stitched into the full image. Later work from \cite{radialDistance}, \cite{Qin2020} and \cite{Zheng2021} tried to design specific complex architecture and loss functions to address the discontinuity issue. In addition, the model performance in pathological cases has been neglected. Structural alterations of the lung, such as honeycombing, bronchial wall thickening and bronchiectasis make the model less robust, leading to local discontinuity and missing bronchi or bronchioles in the segmentation. Variable airway branching patterns between individuals and the data heterogeneity across different medical centres and institutes further complicate the challenge.

To tackle these problems, we present in this paper a simple yet effective adversarial-based patch-scale refinement network to improve the connectivity of preliminary segmentation on both normal and pathological cases. We reason that since it is difficult to derive a single model to accomplish the task, refinement can be done to the existing segmentation. Compared to other similar GAN-based medical segmentation models, we have several novel improvements. Specifically, \textit{tanh} is used as the final activation function in the generator to remove or add pixels to the preliminary labels guided by two centreline objective functions. The refined labels are then dilated using a fixed-size kernel and then times with the CT scan image before feeding into the discriminator. The dilation helps expand the visual field of the discriminator, which we assume could better help it distinguish real and fake pairs. We let the discriminator take image patches instead of the complete image because this allows only penalizing false synthesized structure at the local scale. We evaluate our model on two datasets: the BAS dataset, and our in-house dataset containing 25 cases of both cystic fibrosis and COVID-19, achieving more than 15\% improvement in detected length ratio, detected branch ratio and false negative rate. The refinement pipeline also works on preliminary segmentation of other models, suggested by rises of more than 10\% in the metrics mentioned above. Figures of refined results showed the discontinuity is fixed by our model, making the model more visually interpretable. We also did an ablation study to analyze the contribution of each component in our refinement scheme. 

The main contributions of this paper can be summarised as follows:
\begin{itemize}
    \item Our method provides a new angle for tackling discontinuity in airway segmentation. Instead of delineating the airway right from the start, the refinement model learns to add or remove pixels from the initial segmentation, aiming for better continuity. Moreover, the proposed refinement pipeline can extend to any existing segmentation models.
    \item Our method utilizes patch-scale discriminators to help the generator better focus on tiny structures.
    \item Two novel centreline-based loss functions are implemented to help the generator model maintain the continuity of refined results.
    \item The refinement pipeline is also tested on pathological cases, which were neglected by previous research, and achieves state-of-the-art performance. 
\end{itemize}
\begin{figure*}[!hbt]
    \includegraphics[width=0.75\textwidth]{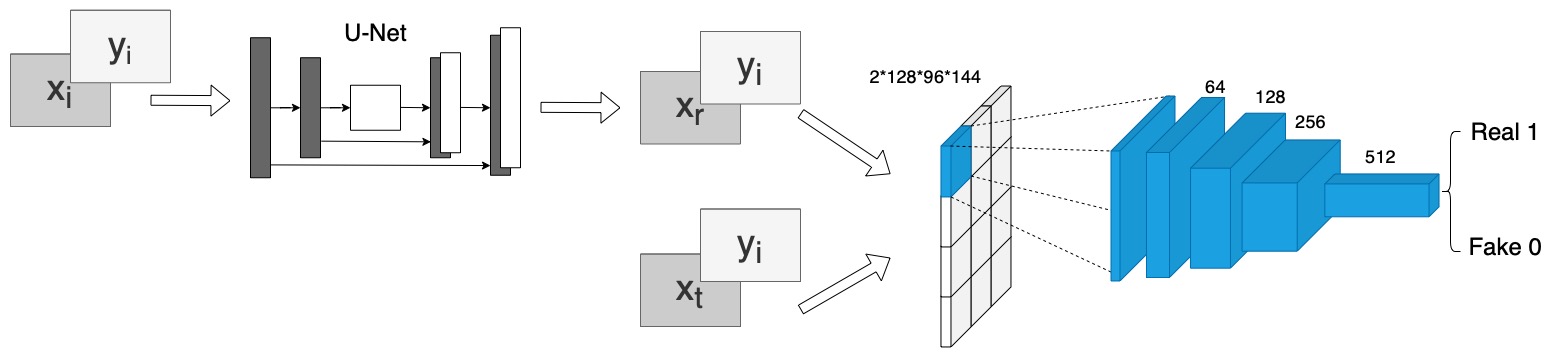}
    \centering
    \caption{The structure of the PatchGAN-based refinement model: U-Net with multi-scale supervision outputs the refined segmentation which is then fed into the PatchGAN discriminator that has five convolution layers. The channel number is denoted on each feature map. }
    \label{fig:patchgan}
\end{figure*}
\section{Related Works}
\label{sec:related_works}
\subsection{Airway Segmentation}
The birth of the U-Net family \cite{Unet}\cite{3dUnet}\cite{Vnet}\cite{nnUnet} shifted the trend in medical image segmentation, and encoding-decoding-based architectures have dominated the field since then. Many deep-learning models have been proposed to segment the human airway structure and some of them have already achieved decent results in terms of IoU, detected length ratio and detected branch ratio. Qin et al. \cite{Qin2020} proposed the two-step AirwayNet to maintain the high performance of the CNN on segmenting peripheral bronchioles. Zheng et al. \cite{Zheng2021} proposed WingsNet with group supervision to enhance the learning of small structures by CNN. They also designed a new objective function termed general union loss to address the imbalance between large and small airways. Charbonnier et al. \cite{CHARBONNIER201752} proposed a ConvNet to detect and remove airway segmentation leakage. They came up with a way to combine segmentation generated by varying parameters to increase the length detected and reduce leakage. Jin et al. \cite{Jin2017} used a 3-D FCN to produce high-quality labels from incomplete ones. They also employed a graph-based refinement incorporating fuzzy connectivity and skeletonisation. Wang et al. \cite{radialDistance} designed a novel radial distance loss function based on distance transform that helps the network recover tiny tubular structures. Nonetheless, the problem of discontinuous predictions and miss detected tiny branches still exists. 

\subsection{Transformer-based medical segmentation}
Transformers were originally proposed for NLP-related tasks due to their ability to encode long-range dependencies, but a recent study \cite{ViT} showed that the architecture can also work with images. Since then, we have witnessed a surge of transformers applications in the medical imaging domain. For 3-D segmentation tasks, Wang et al. \cite{Wang2021TransBTS} proposed TransBTS which positioned a transformer module in between an encoder and a decoder as the bottleneck layer. Hatamizadeh et al. \cite{Hatamizadeh2022} introduced a U-Net Transformer (UNETR) that uses a transformer as the encoding part of the U-Net. Li et al. \cite{Li2021} proposed a Squeeze-and-Expansion Transformer where a squeezed attention block regularizes the self-attention module, and an expansion block learns diversified representations. Liu et al. \cite{Liu2021} proposed a hierarchical transformer which adds a shifted window scheme to the multi-head attention (MSA) module. Unlike our methods, all previous work used the transformer module as a generator, not as a discriminator.
\subsection{Adversarial-based medical segmentation}
GAN-based segmentation has gained a lot of momentum in the field of medical imaging in recent years. Zhang et al. \cite{Zhang2020} used a modified 2-D dense U-Net to generate initial predictions of the airway and then employed a series of morphological operations to extract the small airways. A 3-D dense U-net is then used to learn from the sample of small airways using adversarial training with a cGAN. Zhao et al. \cite{Zhao2020} proposed a generative adversarial learning model with large receptive fields using the dilated residual block as the generator and a multi-layer CNN with residual blocks as the discriminator. Guo et al. \cite{Guo2020} used a dense U-Net with an inception module as the generator and a multi-layer CNN as the discriminator. Park et al. \cite{Park2020} designed an M-GAN which also consists of a multi-layer CNN with residual blocks as the discriminator. Different from our patch-scale approach, despite there being some variations in the structure, previous works in medical segmentation all let the discriminator take in the complete image. Some of them have a complex multi-stage pipeline and didn't achieve competitive results compared with others.
\section{Methodology}
\label{sec:methods}
This section details our novel refinement method, including a generator $\mathcal{G}$ and a discriminator $\mathcal{D}$. The generator takes in the preliminary segmentation $z_i$ along with its corresponding CT image $y_i$ and outputs a refined label $x_r = \mathcal{G}(y_i,x_i)$. The discriminator $\mathcal{D}$ then tries to distinguish the $y_i \cdot (x_r \oplus k)$ from the $y_i \cdot (x_t \oplus k)$, where $k$ is a kernel of size $5\times5\times5$, and $x_t$ is the ground truth.

\begin{figure*}[!hbt]
    \centering
    \includegraphics[width=0.75\textwidth]{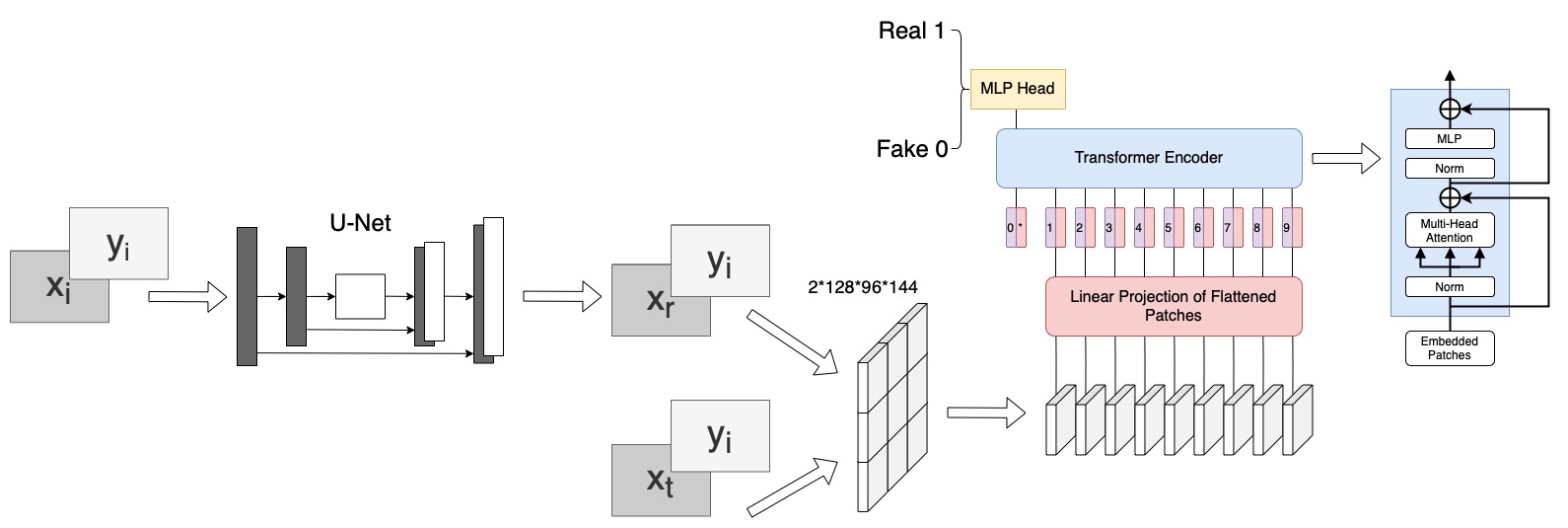}
    \caption{The structure of adversarial ViT-based refinement model: U-Net with multi-scale supervision outputs the refined segmentation which is then fed into the ViT. The ViT splits the image into patches and embeds them into a set of encoded vectors of fixed dimension. The position of each patch is embedded along with the encoded vector.}
    \label{fig:vit}
\end{figure*}

\subsection{Generator}
A 3-D U-Net \cite{Ronneberger2015}\cite{3dUnet} with multi-scale deep supervision is adopted \cite{Lee2014} as the generator to segment the airway structure. CT images and preliminary labelling are combined into a 2-channel input and cut into patches with a dimension of $128\times96\times144$ before feeding into the generator. Preliminary labelling is generated using a small shallow version of U-net \cite{Juarez2018} because it is light and fast to train. The metrics describing the preliminary results are listed in Table \ref{tab:preliminary}. The final activation layer is changed to \textit{tanh} instead of sigmoid since \textit{tanh} produces output that lies in the range $[-1,1]$ and we set 0 as the threshold. For output smaller than zero, it is classified as -1 which aims to remove the false positive region in the preliminary labels; For output larger than zero, it is classified as 1 which aims to remove the false negative region in the preliminary labels. Specific loss functions are used to maintain the connectivity of the refinement which will be discussed in a later section.
\subsection{Discriminator}
Since previous work in medical segmentation has never tried to use discriminators on the scale of patches, we would like to test two types of them. One is the PatchGAN \cite{Isola2016}, which assumes the pixels separated by a patch distance are independent, and it penalizes structure at the scale of parches. The other one is the vision transformer\cite{ViT}, which models an image as a sequence of patches and focuses on the long-range dependencies between them. We hypothesized that if the discriminator can better distinguish the ground truth from the refined labels, then the generator can be better trained to focus on small-scale structures and therefore less breakage in the peripheral bronchioles.
Two different discriminators are tested in our paper. First, A Markovian discriminator is utilized to distinguish a refined label from a true label. Specifically, both refined labels and true labels are dilated using a kernel of size $5\times5\times5$ and then timed with the corresponding CT images prior to feeding into the discriminator, which then tries to classify each $70\times70\times 70$ patch as a synthesized label or ground truth. Second, We adopted a ViT-small(number of layers = 12, Hidden Size = 768, MLP size = 3072, heads = 12) as our discriminator and change the \textit{tanh} in the MLP head to LeakyReLU with negative slope set to 0.2, and specified the patch size to be $32\times38\times36$.
\subsection{Loss Functions}
We use a hybrid loss function to train the refinement model. Specifically, the GAN adversarial loss is incorporated with Dice\cite{Vnet}, cl-Dice\cite{Shit2020} and Continuity and Completeness F-score (CCF) \cite{nan2022} defined as follows: 
\begin{equation}
    \mathrm{Dice} = \frac{2\sum_{i=1}^{N}p_ig_i}{\sum_{i=1}^{N}p_i^2+\sum_{i=1}^{N}g_i^2}
\end{equation}
where $p_i \in P$ is the predicted binary segmentation and $g_i \in G$ is the ground truth binary volume. The summation is taken over $N$ voxels. To maintain the topological integrity of the predicted airway labels, we used clDice defined as:
\begin{equation}
    \mathrm{clDice} = \frac{2\times T_{prec}\times T_{sens}}{T_{prec}+T_{sens}}
\end{equation}
where \emph{Topology precision} $T_{prec} = \frac{| S_P \cap V_L |}{| S_P |}$ and \emph{Topology sensitivity} $T_{sens} = \frac{| S_L \cap V_P |}{| S_L |}$. $V_L$ represents the ground truth mask and $V_P$ represents the predicted mask. $S$ is the skeletonised version of $V$. Dice and clDice are together in the following manner:
\begin{equation}
    L_D = (1-\alpha)(1-Dice)+\alpha (1-clDice)
\end{equation}
where $\alpha \in [0,0.5]$ is a weight parameter.
To further enhance the centreline detection, we employed another objective function focused on continuity and completeness: 
\begin{align}
    \begin{split}
     L_{CCF} & = 1 - C \\
            & = 1 - \frac{\sum X \cdot Y_{CL}}{\sum Y_{CL}}
    \end{split}
\end{align}
where $X$ is the predicted airway labels and $Y_{CL}$ is the centreline of the ground truth. L1 loss is used to maintain the overall segmentation accuracy.
\begin{equation}
    L_{1} = \lvert \sigma(x_i) - y_i \rvert
\end{equation}
The three loss functions described above are combined in the following manner across each layer $j$ of supervision:
\begin{equation}
    L_j = \alpha L_{1} + \beta L_{CCF} + \gamma L_D + \delta L_{cGAN}
\end{equation}
where $\alpha$, $\beta$, $\gamma$ and $\delta$ are weight parameters with range $[0,1.0]$
The total loss over all layers of supervision in the generator is combined with the GAN adversarial loss:
\begin{equation}
    L_{cGAN}(G,D) = \mathbb{E}_{x,y}[logD(x,y)] + \mathbb{E}_{x,z}[logG(x,z)]
\end{equation}
\begin{equation}
    L = \phi_1 L_{layer1} + \phi_2 L_{layer2} + \phi_3 L_{layer3} + \phi_4 L_{final}
\end{equation}
where $\phi_1$ to $\phi_4$ are weight parameters for each layer.
\section{Experiments and Results}
\label{sec:exps}
\subsection{Dataset}
We trained and evaluated our model on the Binary Airway Segmentation (BAS) dataset which contains 90 CT scans with 20 of them from the training set of EXACT'09 \cite{exact09}\footnote{Accessible at http://image.diku.dk/exact/} and 70 of them are from LIDC-IDRI \cite{LIDC}. The original LIDC-IDRI dataset includes 1018 cases but with no airway annotations. \cite{Qin2020} and \cite{Zheng2021} selected 70 cases whose slice thickness is less than or equal to 0.625mm and carefully annotated them\footnote{Accessible at https://geronsushi.github.io/lung.html}. We split the 90 scans into 72 for training and 18 for testing. In addition, we also tested our model on the in-house fibrosis datasets (25 cases) and COVID-19 datasets (25 cases) respectively.  
\subsection{Implementation Details}
The model was implemented using PyTorch and trained on NVIDIA GeForce RTX 3090 for 150 epochs. For the generator part, we used an \textit{Adam} solver with a learning rate = 0.001, and the learning rate was set to decay by half at the 50th, 80th, 100th and 120th epochs. For the discriminator part, we used an \textit{Adam} solver with learning rate = 0.001 betas = 0.5, 0.999. The backpropagation was first performed on the discriminator and then on the generator. Data augmentation was performed on the fly via random horizontal and vertical flips (p=0.5).

During the inference on the test dataset, the input patches are extracted using a sliding window manner with 50\% of overlap in the x,y, and z directions. Patches of binary prediction are stitched back to the full image and then the largest connected component is preserved to reduce noise.
\subsection{Evaluation Metrics}
Given the binary voxel-wise prediction $X$ and the ground truth $Y$, we adopt IoU and Dice coefficient to measure the overall segmentation accuracy as well as other metrics specific for tree-like structures, defined as follows:

\textit{\textbf{IoU and Dice Coefficient}} both measure the proportion of overlapping between the prediction and the ground truth:
\begin{equation}
    IoU = \frac{XY}{X+Y-XY}
\end{equation}
\begin{equation}
    Dice = \frac{2XY}{X+Y} = \frac{2IoU}{IoU + 1}
\end{equation}

\textit{\textbf{Detected Length Ratio (DLR)}} measures the proportion of detected branch length with respect to that of the ground truth:
\begin{equation}
    DLR =  \frac{L_X}{L_Y} 
\end{equation}
where $L_X$ is the total length of the correctly detected airway, $L_Y$ is the total length of the airway in the ground truth.

\textit{\textbf{Detected Branch Ratio (DBR)}} measures the proportion of detected branch number with respect to that of the ground truth:
\begin{equation}
    DBR = \frac{N_X}{N_Y}
\end{equation}
where $N_X$ is the total number of correctly detected airway branches, $N_Y$ is the number of branches in the ground truth. In this study, branches with the intersection over union (IoU) score greater than 0.8 are referred to be correctly identified.

\textit{\textbf{Precision}} refers to the fraction of correctly identified airway voxels among the predicted airway:
\begin{equation}
    Precision = \frac{TP}{TP+FP}
\end{equation}
where $TP$ and $FP$ are the numbers of true positive voxels and false positive voxels.

\textit{\textbf{Leakage}} measures the proportion of total false positives with respect to the ground truth annotations:
\begin{equation}
    Leakage = \frac{V_X}{V_Y}
\end{equation}
where $V_X$ is the volume of false-positive predictions, $V_Y$ is the volume of ground truth annotations.

\textit{\textbf{Airway Missing Ratio (AMR)}} measures the proportion of total undetected airways (false negatives) with respect to the ground truth annotations:
\begin{equation}
    AMR = \frac{FN}{V_Y}
\end{equation}
where $FN$ is the volume of false-negative predictions, $V_Y$ is the volume of ground truth annotations.
\subsection{Segmentation Results}
\begin{figure}
    \centering
    \includegraphics[width=0.45\textwidth]{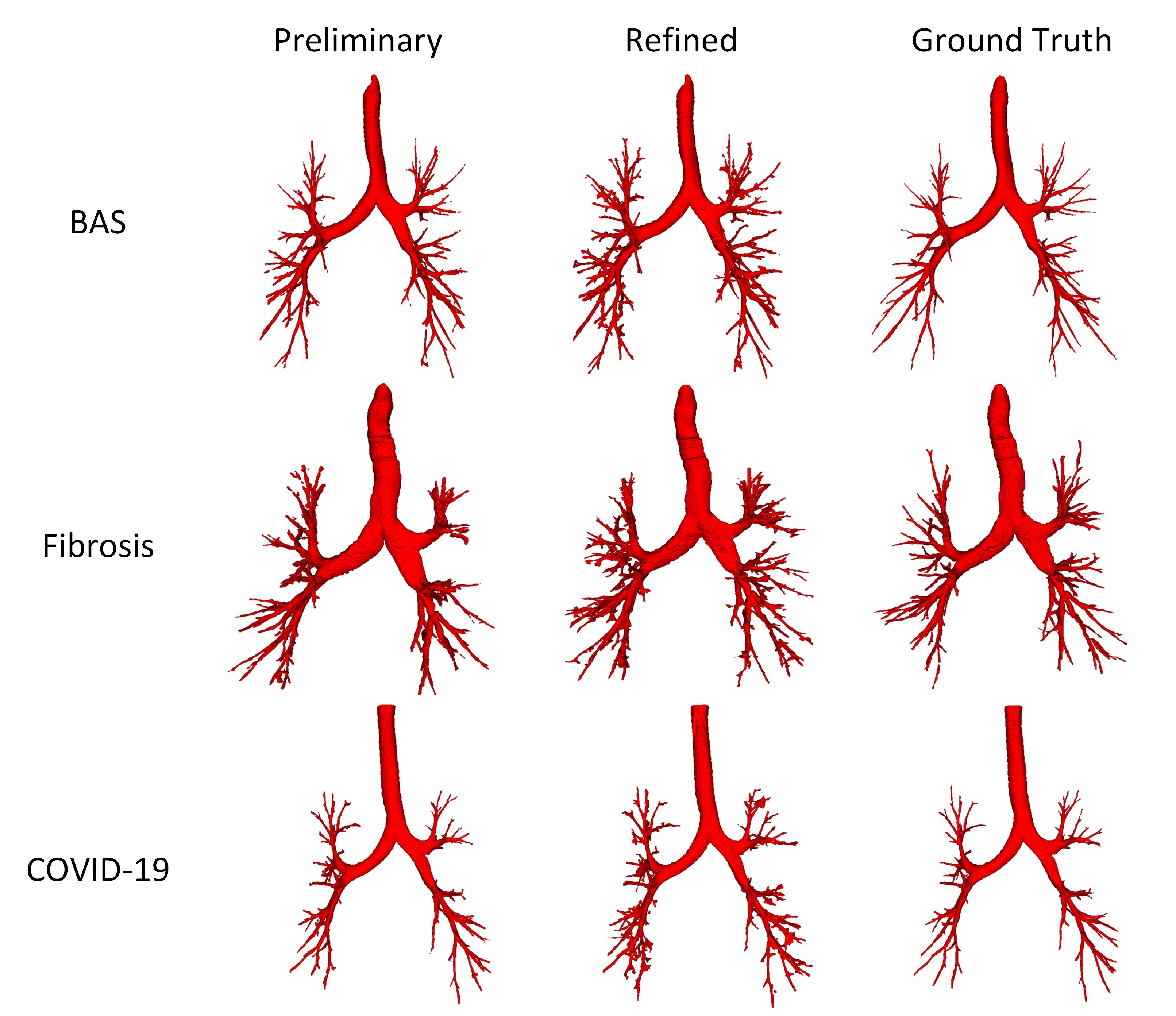}
    \caption{Visualization of preliminary segmentation, refined segmentation and the ground truth on BAS (CASE02) and our in-house dataset ($162\_01$ in fibrosis and RM451 in COVID-19)}
    \label{fig:results_combined}
\end{figure}

\begin{figure}
    \centering
    \includegraphics[width=0.5\textwidth]{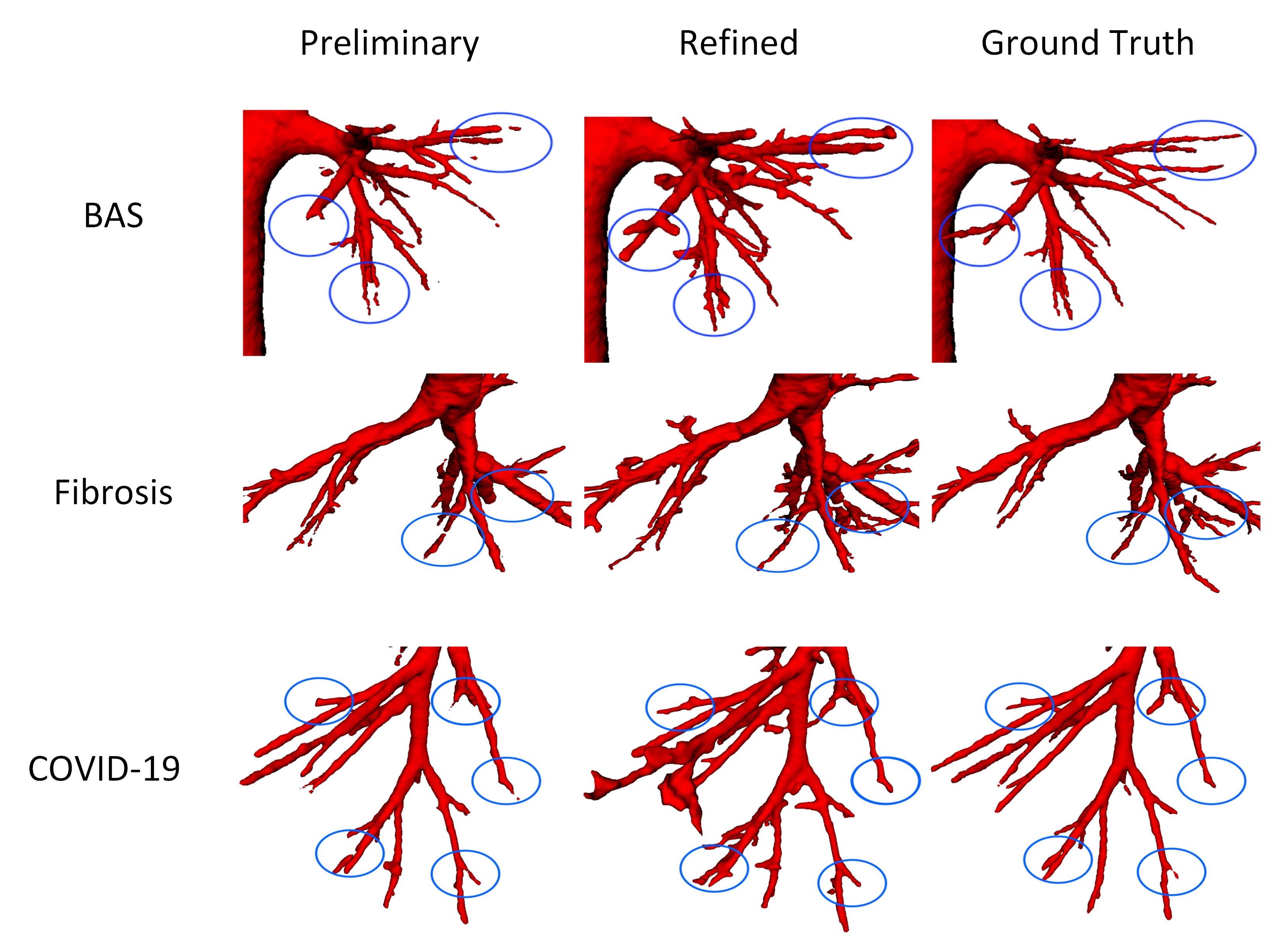}
    \caption{Visualization of the zoomed-in region in preliminary segmentation, refined segmentation and the ground truth on BAS (CASE02) and our in-house dataset ($162\_01$ in fibrosis and RM451 in COVID-19)}
    \label{fig:results_zoom_combined}
\end{figure}

We use seven evaluation metrics including intersection over union (IoU), dice coefficient, detected length ratio (DLR), detected branch ratio (DBR), precision, leakage and false negative rates (AMR) to evaluate the performance of our model comprehensively. Furthermore, we perform Wilcoxon signed-rank test $(\alpha = 0.01)$.for statistical analysis. 
\begin{table*}[!htbp]
\center
\caption{Preliminary segmentation results}
\label{tab:preliminary}
\resizebox{\textwidth}{!}{%
\begin{tabular}{lllllllllllll}
\toprule
& Model                                                      & IoU                            & Dice                           & DLR                    & DBR                     & Precision                      & Leakages                       & AMR &   \\ \midrule
Valid & small U-Net                              & 0.8372±0.0752          & 0.9093±0.0505           & 0.7346±0.1419           & 0.6425±0.1555            & 0.9344±0.0314          & 0.0621±0.0269           & 0.1117±0.0762 &  \\ \midrule
Fibrosis & small U-Net & 0.7911±0.0532 & 0.8823±0.0349 & 0.5616±0.1076 & 0.4829±0.1218 & 0.9296±0.0294 & 0.0657±0.0362 & 0.1568±0.0637
&  
\\ \midrule    
COVID & small U-Net & 0.8637±0.10165 & 0.9228±0.0756 & 0.7173±0.1507 & 0.6302±0.1507 & 0.9566±0.0134 & 0.0407±0.0137 & 0.1008±0.1079 &  
\\ \bottomrule
\end{tabular}%
}
\end{table*}
\noindent
\textbf{Comparison on BAS dataset:}
Our proposed method achieved state-of-the-art performance on BAS (Table \ref{tab:results} and Figure \ref{fig:results_combined}, \ref{fig:results_zoom_combined}) with 0.8902 DLR, 0.8439 DBR and 0.05441 AMR. WingsNet achieves the highest IoU (0.8544) and voxel-wise precision (0.9458) among all models. NaviAirway has the lowest AMR (0.0413) while the model proposed by Wang et al. suffers from high leakage (0.3070). NaviAirway and the model proposed by Wang et al. also have competitive performance in DLR and DBR. 

\noindent
\textbf{Comparison on in-house fibrosis dataset:}
All models performance decreases drastically in the fibrosis cases (Table \ref{tab:results} and Figure \ref{fig:results_combined}, \ref{fig:results_zoom_combined}). Notwithstanding, our model still achieves the highest DLR (0.7242), DBR (0.6550) and lowest AMR (0.08232). WingsNet and the model proposed by Wang et al. achieve similar results on DLR and DBR, but WingsNet has a better IoU score. NaviAirway obtains the highest IoU while unexpectedly achieving a poor performance on maintaining continuity with 0.5994 DLR and 0.5148 DBR.

\noindent
\textbf{Comparison on in-house COVID-19 dataset:}
All models perform slightly better compared to them on the BAS dataset (Table \ref{tab:results} and Figure \ref{fig:results_combined}, \ref{fig:results_zoom_combined}). The proposed method achieves the best DLR (0.9104), DBR (0.8772) and AMR (0.0281) in the COVID-19 cases. WingsNet achieves the second highest DLR (0.9073) and DBR (0.8682) followed by NaviAirway with 0.9394 DLR and 0.8729 DBR. WingsNet also achieves highest IoU (0.9176), Dice (0.9566) and precision (0.9683)

\begin{figure}[!htbp]
     \centering
     \includegraphics[width=0.5\textwidth]{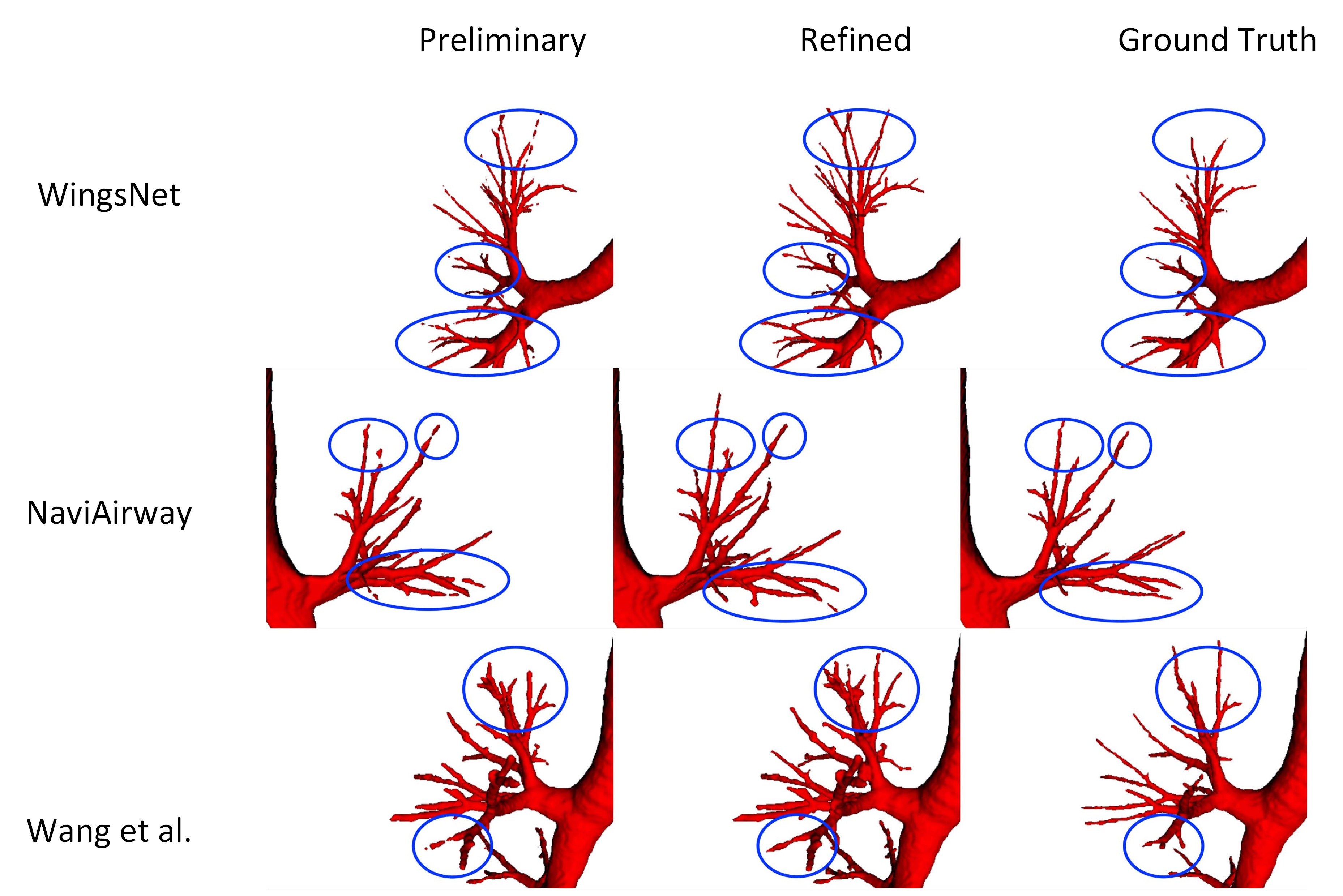}
     \caption{Zoom in on bronchioles of CASE02 in EXACT'09 test set; From top to bottom row are results from WingsNet, NaviAirway and Wang et al. respectively; From left to right column are initial segmentation of their models, refined segmentation and ground truth. Blue circles indicate breakages in the preliminary segmentation, but are later fixed by the refinement model}
     \label{fig:valid_comparison}
\end{figure}

\begin{figure}[!htbp]
     \centering
     \includegraphics[width=0.5\textwidth]{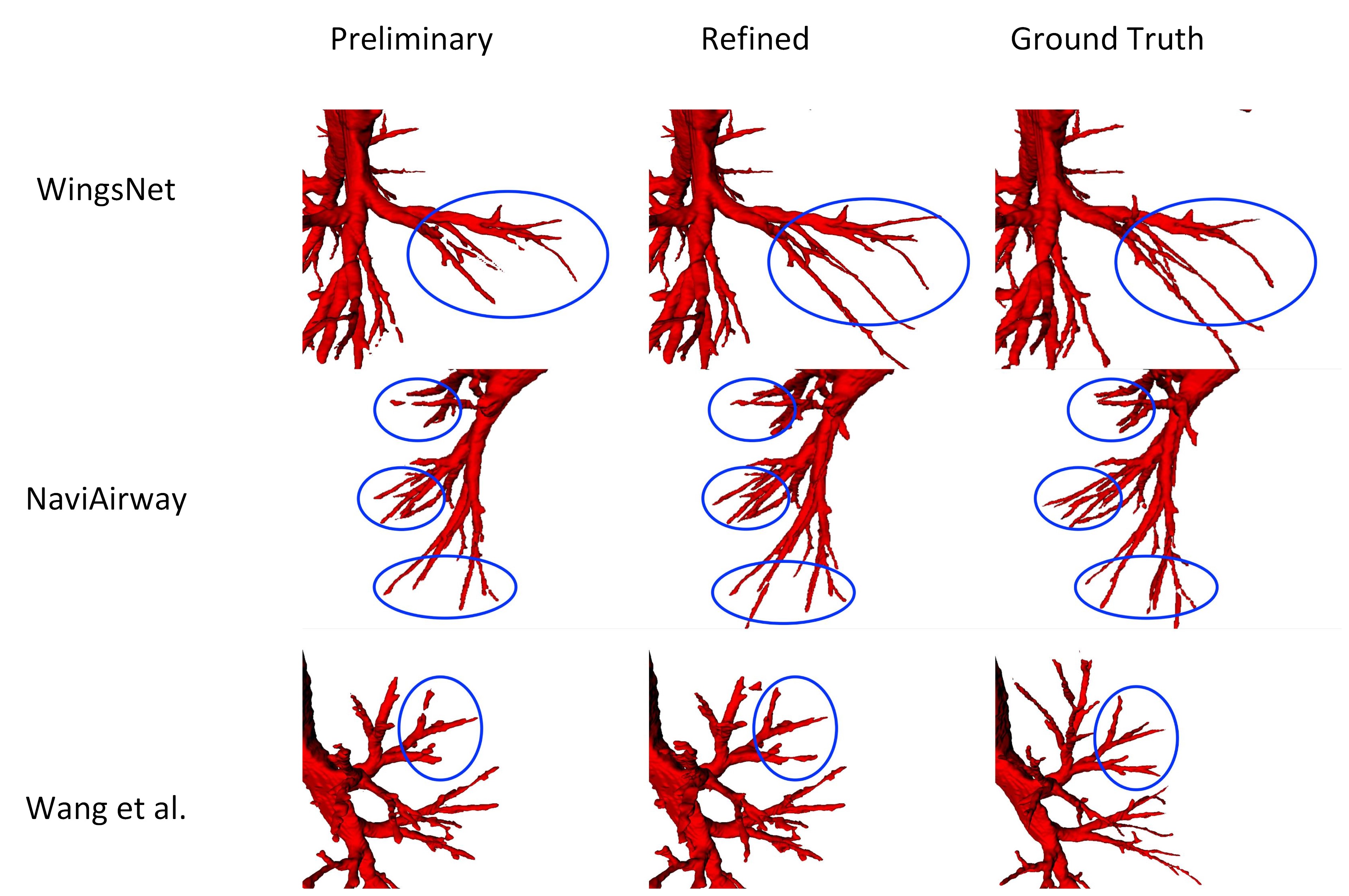}
     \caption{Zoom in on bronchioles of Case $162\_01$ in fibrosis test set; From top to bottom row are results from WingsNet, NaviAirway and Wang et al. respectively; From left to right column are initial segmentation of their models, refined segmentation and ground truth. Blue circles indicate breakages in the preliminary segmentation, but are later fixed by the refinement model}
     \label{fig:fibrosis_comparison}
\end{figure}

\begin{figure}[!htbp]
     \centering
     \includegraphics[width=0.5\textwidth]{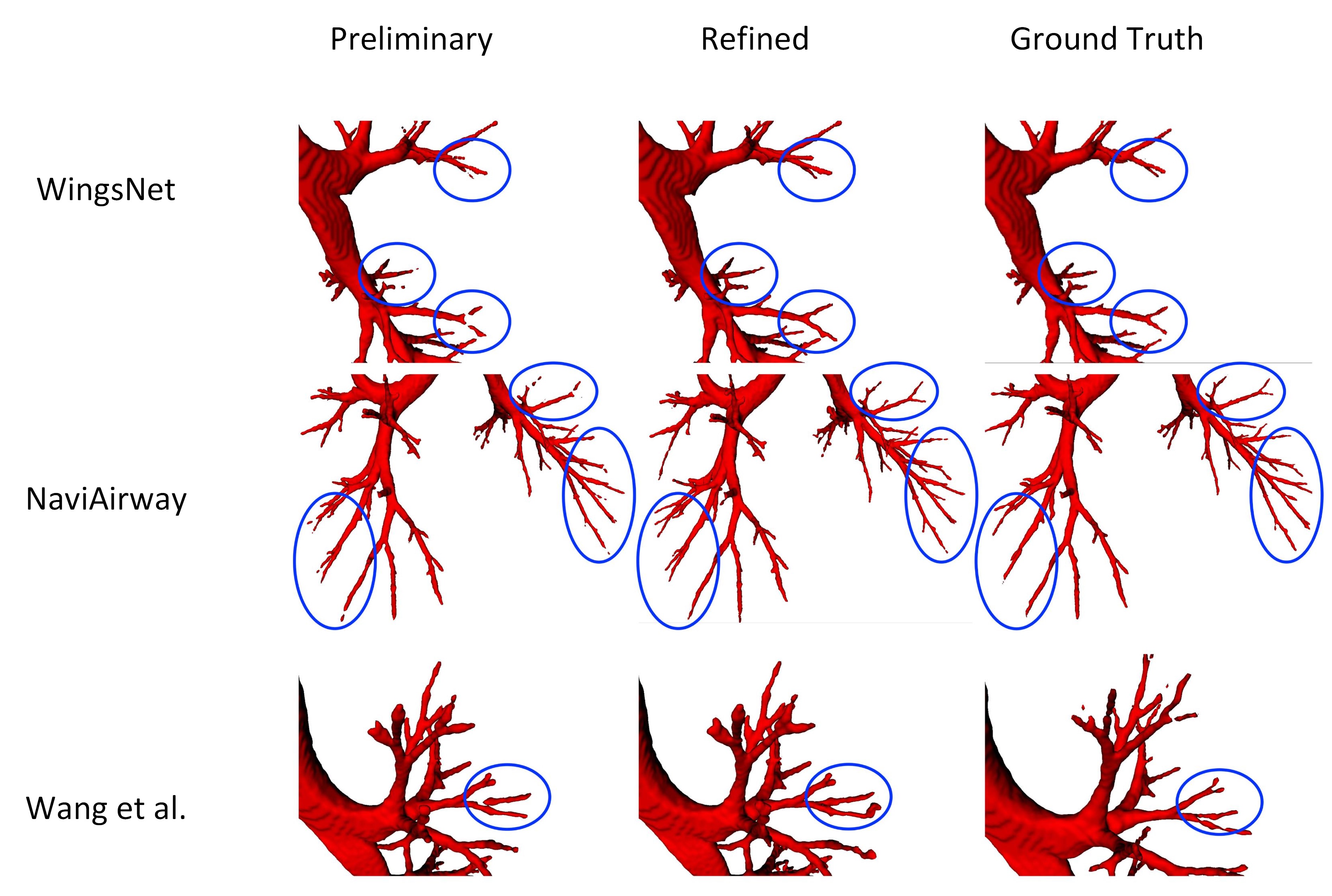}
     \caption{Zoom in on bronchioles of Case RM451 in the COVID test set; From top to bottom row are results from WingsNet, NaviAirway and Wang et al. respectively; From left to right column are initial segmentation of their models, refined segmentation and ground truth. Blue circles indicate breakages in the preliminary segmentation, but are later fixed by the refinement model}
     \label{fig:covid_comparison}
\end{figure}

\begin{table*}[!htbp]
\caption{Comparison experiments on different datasets}
\label{tab:results}
\resizebox{\textwidth}{!}{%
\begin{threeparttable}
\begin{tabular}{lllllllllllll}
\toprule
         & Model         & IoU$\downarrow$                     & Dice$\downarrow$                    & DLR$\uparrow$                     & DBR$\uparrow$                     & Precision$\downarrow$               & Leakage$\uparrow$                 & AMR$\downarrow$                   &   \\ \midrule
Valid   & Wang et al. \cite{radialDistance}\tnote{$\dagger$} & 0.7330±0.0786\tnote{$\ddagger$} & 0.8434±0.0554\tnote{$\ddagger$}  & 0.8505±0.1227\tnote{$\ddagger$} & 0.7858±0.1420\tnote{$\ddagger$} & 0.7636±0.0737\tnote{$\ddagger$} & 0.3070±0.1412\tnote{$\ddagger$}
 & 0.0507±0.0619 &   \\ & WingsNet \cite{Zheng2021}\tnote{*}& \textbf{0.8544±0.0673}\tnote{$\ddagger$} & \textbf{0.9120±0.0422}\tnote{$\ddagger$} & 0.8698±0.1175\tnote{$\ddagger$} & 0.8166±0.1305\tnote{$\ddagger$} & \textbf{0.9458±0.0271}\tnote{$\ddagger$} & \textbf{0.0529±0.0300}\tnote{$\ddagger$} & 0.1002±0.0769\tnote{$\ddagger$} &   \\& NaviAirway \cite{NaviAirway}\tnote{*}& 0.8348±0.0335 & 0.9096±0.0200 & 0.8734±0.0715 & 0.8099±0.0950\tnote{$\ddagger$} & 0.8672±0.0406 & 0.1500±0.0536
 & \textbf{0.0413±0.0304} & \\& \textbf{Unet+PatchGAN} & 0.8150±0.0519  & 0.8971±0.0330 & 0.8828±0.1012    & 0.8337±0.1263    & 0.8556±0.0393   & 0.1620±0.0522   & 0.0544±0.0525     & \\
         & \textbf{Unet+ViT}      & 0.8132±0.0518 & 0.8961±0.0334 & \textbf{0.8902±0.0967}  & \textbf{0.8439±0.1261} & 0.8600±0.0401 & 0.1549±0.0539  & 0.0623±0.0504 & \\ \midrule
Fibrosis & Wang et al. \cite{radialDistance}\tnote{$\dagger$} & 0.6979±0.0647\tnote{$\ddagger$}  & 0.8203±0.0462\tnote{$\ddagger$} & 0.6961±0.0924\tnote{$\ddagger$} & 0.6261±0.1117\tnote{$\ddagger$} & 0.7468±0.0773\tnote{$\ddagger$} & 0.3272±0.1445\tnote{$\ddagger$} & 0.0823±0.0388 &  \\ & WingsNet \cite{Zheng2021}\tnote{*}& 0.8052±0.0539\tnote{$\ddagger$} & 0.8910±0.0440\tnote{$\ddagger$} & 0.6951±0.0977\tnote{$\ddagger$} & 0.6198±0.1175\tnote{$\ddagger$} & \textbf{0.9505±0.0116}\tnote{$\ddagger$} & \textbf{0.0438±0.0112}\tnote{$\ddagger$} & 0.1595±0.0576\tnote{$\ddagger$} & \\& NaviAirway \cite{NaviAirway}\tnote{*}& \textbf{0.8074±0.0533}\tnote{$\ddagger$} & \textbf{0.8924±0.0440}\tnote{$\ddagger$} & 0.5994±0.1440\tnote{$\ddagger$} & 0.5148±0.1490\tnote{$\ddagger$} & 0.9247±0.0165\tnote{$\ddagger$} & 0.0714±0.0188\tnote{$\ddagger$} & 0.1345±0.0645\tnote{$\ddagger$} &   \\& \textbf{Unet+PatchGAN} & 0.7481±0.0678 & 0.8541±0.0464 & 0.7157±0.1067 & 0.6437±0.1195 & 0.8035±0.0775 & 0.2374±0.1286 & \textbf{0.0823±0.0332} & \\
         & \textbf{Unet+ViT}      & 0.7272±0.0631   & 0.8405±0.0436  & \textbf{0.7242±0.1096}   & \textbf{0.6550±0.1266}   & 0.7879±0.0816   & 0.2606±0.1401    & 0.0916±0.0325   & \\ \midrule
COVID  & Wang et al. \cite{radialDistance}\tnote{$\dagger$} & 0.7433±0.1010\tnote{$\ddagger$} & 0.8481±0.0798\tnote{$\ddagger$} & 0.8487±0.1320\tnote{$\ddagger$} & 0.7993±0.1409\tnote{$\ddagger$} & 0.7749±0.0736\tnote{$\ddagger$} & 0.2843±0.1500\tnote{$\ddagger$} & 0.0541±0.1035\tnote{$\ddagger$} & \\ & WingsNet \cite{Zheng2021}\tnote{*} & \textbf{0.9176±0.0363}\tnote{$\ddagger$} & \textbf{0.9566±0.0209}\tnote{$\ddagger$} & 0.9073±0.0647 & 0.8682±0.0831 & \textbf{0.9683±0.0204}\tnote{$\ddagger$} & \textbf{0.0311±0.0203}\tnote{$\ddagger$} & 0.0544±0.0283\tnote{$\ddagger$} &\\& NaviAirway \cite{NaviAirway}\tnote{*} & 0.8861±0.0298\tnote{$\ddagger$} & 0.9394±0.0171\tnote{$\ddagger$} & 0.8729±0.0792\tnote{$\ddagger$} & 0.8060±0.1034\tnote{$\ddagger$} & 0.9100±0.0245\tnote{$\ddagger$} & 0.0970±0.0316\tnote{$\ddagger$} & 0.0286±0.0239 &  \\& \textbf{Unet+PatchGAN} & 0.8292±0.0335 & 0.9063±0.0203 & 0.8911±0.0827 & 0.8443±0.1044 & 0.8506±0.0413 & 0.1741±0.0575    & \textbf{0.0281±0.0258}    & \\
         & \textbf{Unet+ViT}      & 0.8090±0.0483   & 0.8936±0.0307  & \textbf{0.9104±0.0751}   & \textbf{0.8772±0.0987}   & 0.8354±0.0458   & 0.1933±0.0705   & 0.0378±0.0231   &  \\ \bottomrule
         
\end{tabular}
\begin{tablenotes}
\item[*] refers to results obtained from open-source implementations with model weights provided.
\item[$\dagger$] refers to reproduced results.
\item[$\ddagger$] represents statistical significance (with Wilcoxon signed-rank test $p < 0.01$) compared with the proposed method.
\end{tablenotes}
\end{threeparttable}%
}
\end{table*}
\noindent
\textbf{Refined segmentation on other three models:}
To demonstrate our refinement pipeline can be extended to other models, we refined all three other aforementioned models (WingsNet, NaviAirway and the model proposed by Wang et al.). Overall, all three models' performance on all three datasets (BAS, fibrosis and COVID-19) improved significantly with DLR and DBR increasing more than 10\% on average (Table \ref{tab:refined_seg}). The IoU and Dice decrease a little as a trade-off in all scenarios with some even not considered to be statistically significant. The false negative ratio is also reduced significantly in all cases. Notably, the improvement on these three models is not as much as the refinement of our own preliminary results. Figure \ref{fig:valid_comparison} to \ref{fig:covid_comparison} provide a zoom-in illustration of the effect of our refinement visually. 

\begin{table*}[!htbp]
\caption{Refined segmentation results}
\label{tab:refined_seg}
\resizebox{\textwidth}{!}{%
\begin{threeparttable}
\begin{tabular}{lllllllllllll}
\toprule
& Model                                                      & IoU$\downarrow$                            & Dice$\downarrow$                           & DLR$\uparrow$                    & DBR$\uparrow$                     & Precision$\downarrow$                      & Leakages$\uparrow$                       & AMR$\downarrow$ &\\ \midrule
Valid & Wang et al. \cite{radialDistance}{$\dagger$} & 0.7330±0.0786\tnote{$\ddagger$} & 0.84348±0.0554\tnote{$\ddagger$}  & 0.8505±0.1227\tnote{$\ddagger$} & 0.7858±0.1420\tnote{$\ddagger$} & 0.7636±0.0737\tnote{$\ddagger$} & 0.3070±0.1412\tnote{$\ddagger$}
 & 0.0507±0.0619\tnote{$\ddagger$} &  \\& Refined Wang et al. & 0.7136±0.1000 & 0.8284±0.0763 & 0.9093±0.1263 & 0.8700±0.1380 & 0.7341±0.0793 & 0.3552±0.1281 & 0.0423±0.0960 &  \\\cmidrule{2-9}& WingsNet \cite{Zheng2021}\tnote{*} & 0.8544±0.0673 & 0.9200±0.0422 & 0.8698±0.1175\tnote{$\ddagger$} & 0.8166±0.1305\tnote{$\ddagger$} & 0.9458±0.0271\tnote{$\ddagger$} & 0.0529±0.0300\tnote{$\ddagger$} & 0.1002±0.0769\tnote{$\ddagger$} &  \\ & Refined WingsNet & 0.8504±0.0651 & 0.9177±0.0409 & 0.9050±0.1032 & 0.8621±0.1230 & 0.9331±0.0303 & 0.0668±0.0347 & 0.0927±0.0752 & \\\cmidrule{2-9} & NaviAirway \cite{NaviAirway}\tnote{*}& 0.8348±0.0335\tnote{$\ddagger$} & 0.9096±0.0200\tnote{$\ddagger$} & 0.8734±0.0715\tnote{$\ddagger$} & 0.8099±0.0950\tnote{$\ddagger$} & 0.8672±0.0406\tnote{$\ddagger$} & 0.1500±0.0536\tnote{$\ddagger$}
 & 0.0413±0.0304\tnote{$\ddagger$} & \\ & Refined NaviAirway & 0.8246±0.0354 & 0.9034±0.0213 & 0.9064±0.0595 & 0.8590±0.0857 & 0.8516±0.0432 & 0.1716±0.0584 & 0.0357±0.0261 & \\ \midrule
Fibrosis & Wang et al. \cite{radialDistance}\tnote{$\dagger$} & 0.6979±0.0647\tnote{$\ddagger$}  & 0.8203±0.0462\tnote{$\ddagger$} & 0.6961±0.0924\tnote{$\ddagger$} & 0.6261±0.1117\tnote{$\ddagger$} & 0.7468±0.0773\tnote{$\ddagger$} & 0.32727±0.1445\tnote{$\ddagger$} & 0.0822±0.0388 &  \\& Refined Wang et al. & 0.6528±0.0925 & 0.7859±0.0722 & 0.8007±0.0847 & 0.7399±0.1042 & 0.6736±0.0951 & 0.4953±0.2456 & 0.0456±0.0277 &\\ \cmidrule{2-9} & WingsNet \cite{Zheng2021}\tnote{*}& 0.8052±0.0539\tnote{$\ddagger$} & 0.8910±0.0340\tnote{$\ddagger$} & 0.6951±0.0977\tnote{$\ddagger$} & 0.6198±0.1175\tnote{$\ddagger$} & 0.9505±0.0116\tnote{$\ddagger$} & 0.0438±0.0113\tnote{$\ddagger$} & 0.1595±0.0576\tnote{$\ddagger$} & \\ & Refined WingsNet & 0.8035±0.0515 & 0.8901±0.0326 & 0.7318±0.1012\tnote{$\ddagger$} & 0.6638±0.1221\tnote{$\ddagger$} & 0.9341±0.0159 & 0.0604±0.0163 & 0.1480±0.0565 &  \\ \cmidrule{2-9} & NaviAirway \cite{NaviAirway}\tnote{*}& 0.8074±0.0533\tnote{$\ddagger$} & 0.8924±0.0340\tnote{$\ddagger$} & 0.5994±0.1440\tnote{$\ddagger$} & 0.5148±0.1490\tnote{$\ddagger$} & 0.9247±0.0165\tnote{$\ddagger$} & 0.0714±0.0188\tnote{$\ddagger$} & 0.1345±0.0645\tnote{$\ddagger$} & \\& Refined NaviAirway & 0.8061±0.0514 & 0.8917±0.0329 & 0.6534±0.1513 & 0.5725±0.1617 & 0.9090±0.0189 & 0.0891±0.0226 & 0.1215±0.0657 & \\ \midrule 
COVID & Wang et al. \cite{radialDistance}{$\dagger$} & 0.7433±0.1010 & 0.8481±0.0798 & 0.8487±0.1320\tnote{$\ddagger$} & 0.7993±0.1409\tnote{$\ddagger$} & 0.7749±0.0736\tnote{$\ddagger$} & 0.2843±0.1500\tnote{$\ddagger$} & 0.0541±0.1035\tnote{$\ddagger$} & \\& Refined Wang et al. & 0.7408±0.0566 & 0.8498±0.0392 & 0.9538±0.0435 & 0.9316±0.0603 & 0.7462±0.0562 & 0.3453±0.1152 & 0.0098±0.0106 &  \\ \cmidrule{2-9} & WingsNet \cite{Zheng2021}\tnote{*} & 0.9176±0.0363\tnote{$\ddagger$} & 0.9566±0.0209\tnote{$\ddagger$} & 0.9073±0.0647\tnote{$\ddagger$} & 0.8682±0.0831\tnote{$\ddagger$} & 0.9683±0.0204\tnote{$\ddagger$} & 0.0311±0.0203\tnote{$\ddagger$} & 0.0544±0.0283{$\ddagger$} & \\  & Refined WingsNet & 0.8971±0.0345 & 0.9454±0.0200 & 0.9571±0.0575 & 0.9409±0.0778 & 0.9424±0.0266 & 0.0588±0.0297 & 0.0510±0.0259 &\\ \cmidrule{2-9} & NaviAirway \cite{NaviAirway}\tnote{*} & 0.8861±0.0298\tnote{$\ddagger$} & 0.9394±0.0171\tnote{$\ddagger$} & 0.8729±0.0792\tnote{$\ddagger$} & 0.8060±0.1034\tnote{$\ddagger$} & 0.9100±0.0245\tnote{$\ddagger$} & 0.0970±0.0316\tnote{$\ddagger$} & 0.0286±0.0239\tnote{$\ddagger$} &  \\& Refined NaviAirway & 0.8725±0.0315 & 0.9316±0.0186 & 0.9153±0.0724 & 0.8732±0.0997 & 0.8904±0.0295 & 0.1218±0.0407 & 0.0223±0.0214 & \\\bottomrule
\end{tabular}
\begin{tablenotes}
\item[*] refers to results obtained from open-source implementations with model weights provided.
\item[$\dagger$] refers to reproduced results.
\item[$\ddagger$] represents statistical significance (with Wilcoxon signed-rank test $p < 0.01$) compared with the refined results.
\end{tablenotes}
\end{threeparttable}%
}
\end{table*}
\subsection{Ablation Study}
We run ablation studies to isolate the effect of each component in our refinement model as shown in Table \ref{tab:ablation} including 3-D U-Net + PatchGAN with dilation (BL), BL + clDIce, BL + CCF, BL + CCF with multi-scale supervision, BL+clDice with multi-scale supervision, BL+clDice+CCF with multi-scale supervision and BL (PatchGAN is replaced with ViT)+clDice+CCF with multi-scale supervision. By adopting cl-Dice, all seven metrics are slightly improved (roughly 1.0\% average gain). When clDice is replaced with CCF, IoU, Dice and leakage become worse while the rest four metrics are improved drastically by more than 5.0\% on average. The incorporation of multi-scale supervision helps the model guided by CCF and cl-Dice promote in all seven metrics. Finally, the best performance in DLR (0.8902) and DBR (0.8439) is achieved when CCF, cl-Dice and multi-scale supervision are combined with ViT.
\begin{table*}[!htb]
\caption{Ablation study of the proposed model.}
\label{tab:ablation}
\resizebox{\textwidth}{!}{%
\begin{tabular}{llllllll}
\toprule
Model                                                      & IoU                            & Dice                           & DLR                    & DBR                     & Precision                      & Leakages                       & AMR           \\ \midrule
U-Net+PatchGAN+dilation(BL)                             & 0.8274±0.0563          & 0.9044±0.0355         & 0.7936±0.1303           & 0.7088±0.1557           & 0.8866±0.0405          & 0.1215±0.0504           & 0.0730±0.0622          \\
BL+clDice                     & 0.8325±0.0465          & 0.9079±0.0294          & 0.8059±0.1285           & 0.7324±0.1525            & 0.8895±0.0313          & 0.1178±0.0400          & 0.0693±0.0608          \\
BL+ccf                          & 0.7133±0.0854         & 0.8295±0.0631          & 0.8526±0.1148           & 0.7947±0.1474            & 0.7441±0.0880          & 0.3519±0.2151          & \textbf{0.0516±0.0541} \\
BL+ccf+multi-scale               & 0.7701±0.0522          & 0.8691±0.0343          & 0.8420±0.1277          & 0.7788±0.1542             & 0.8101±0.0484          & 0.2266±0.0761         & 0.0573±0.0600          \\
BL+clDice+multi-scale          & \textbf{0.8342±0.0670} & \textbf{0.9080±0.0427} & 0.8352±0.1156           & 0.7671±0.1421            & \textbf{0.8969±0.0478} & \textbf{0.1082±0.0578} & 0.0781±0.0565          \\
BL+ccf+ clDice + multi-scale   & 0.8150±0.0519           & 0.8971±0.0330         & 0.8828±0.1012            & 0.8337±0.1263             & 0.8556±0.0393          & 0.1620±0.0522          & 0.0544±0.0525            \\
BL(ViT) + ccf + cl\_dice + multi-scale & 0.8133±0.0518        & 0.8961±0.0334        & \textbf{0.8902±0.0967} & \textbf{0.8439±0.1261} & 0.8600±0.0401        & 0.1549±0.0539         & 0.0623±0.0505\\ \bottomrule       
\end{tabular}%
}
\end{table*}

\section{Discussion and Conclusion}
\label{sec:discussion}
In this paper, we proposed a novel adversarial-based refinement model using \textit{tanh} as the final activation function and trained the network using objective functions clDice and CCF that focus on the continuity of the airway. The refinement model corrects breakage and adds missing branches in the preliminary segmentation generated by other networks such as U-Net.

We evaluated our model on two datasets: the BAS dataset, and our in-house dataset containing 25 cases of cystic fibrosis and 25 cases of COVID-19. By comparing preliminary results in Table \ref{tab:preliminary} and refined results in Table \ref{tab:results}, we can find that our method significantly reduces false negatives and increase the length and number of detected branches in all three test datasets. In Figures \ref{fig:results_combined} to \ref{fig:covid_comparison}, we use blue circles to indicate breakage and bronchioles missed in preliminary segmentation but later successfully detected in the refined results. Airway segmentation completeness is essential for the clinical implementation of the algorithm since it can provide biomarkers for evaluating the severity of lung diseases. For example, traction bronchiectasis \cite{Walsh2012}\cite{Walsh2014} and airway tapering \cite{Kuo2020} are useful biomarkers in the prognosis of cystic fibrosis. The ability of our model to maintain the local connectivity and thereby maximize the completeness is desirable for clinical use since methods to evaluate the airway assume disconnected branches are removed when retrieving the largest connected component.


There are also limitations in our current approach. First, the model is trained on a specific set of preliminary segmentation generated using a small U-Net. Therefore, The improvement in the preliminary segmentation produced by the same network (small U-Net) is higher than in the other three models (Wang et al., WingsNet and NaviAirway). Its ability to refine is also constrained by the quality of the preliminary results, which is also suggested in Table. \ref{tab:refined_seg} that better initial results generally lead to better refinement. To solve this issue in the future, we will probably train the model using preliminary segmentation generated by randomly masking out ground truth. Second, although airway continuity is maintained, some peripheral bronchioles are over-segmented after repairing which is also reflected by the higher leakage. The leakage is more recognizable in the fibrosis dataset than it is in the BAS and COVID-19 datasets. Another interesting observation we noticed from figures \ref{fig:results_combined} to \ref{fig:covid_comparison} is that sometimes the predictions have detected correct airway branches not shown in the ground truth. This also could explain why the IoU and Dice decrease a little while leakage rises after refinement. 

In conclusion, we have demonstrated that patch-scale discriminators can help improve airway segmentation in terms of better connectivity and lower false negative rates. This refinement pipeline can also be extended to other models and segmentation tasks in the future.


%

\appendices


\ifCLASSOPTIONcaptionsoff
  \newpage
\fi



%

\bibliographystyle{IEEEtran}
 \bibliography{cas-refs}

\begin{thebibliography}{10}
\providecommand{\url}[1]{#1}
\csname url@samestyle\endcsname
\providecommand{\newblock}{\relax}
\providecommand{\bibinfo}[2]{#2}
\providecommand{\BIBentrySTDinterwordspacing}{\spaceskip=0pt\relax}
\providecommand{\BIBentryALTinterwordstretchfactor}{4}
\providecommand{\BIBentryALTinterwordspacing}{\spaceskip=\fontdimen2\font plus
\BIBentryALTinterwordstretchfactor\fontdimen3\font minus
  \fontdimen4\font\relax}
\providecommand{\BIBforeignlanguage}[2]{{%
\expandafter\ifx\csname l@#1\endcsname\relax
\typeout{** WARNING: IEEEtran.bst: No hyphenation pattern has been}%
\typeout{** loaded for the language `#1'. Using the pattern for}%
\typeout{** the default language instead.}%
\else
\language=\csname l@#1\endcsname
\fi
#2}}
\providecommand{\BIBdecl}{\relax}
\BIBdecl

\bibitem{thresholding}
D.~Aykac, E.~Hoffman \emph{et~al.}, ``Segmentation and analysis of the human
  airway tree from three-dimensional x-ray ct images,'' \emph{IEEE Transactions
  on Medical Imaging}, vol.~22, no.~8, pp. 940--950, 2003.

\bibitem{region_growing01}
J.~Tschirren, E.~Hoffman \emph{et~al.}, ``Intrathoracic airway trees:
  segmentation and airway morphology analysis from low-dose ct scans,''
  \emph{IEEE Transactions on Medical Imaging}, vol.~24, no.~12, pp. 1529--1539,
  2005.

\bibitem{region_growing02}
M.~W. Graham, J.~D. Gibbs \emph{et~al.}, ``Robust 3-d airway tree segmentation
  for image-guided peripheral bronchoscopy,'' \emph{IEEE Transactions on
  Medical Imaging}, vol.~29, no.~4, pp. 982--997, 2010.

\bibitem{exact09}
P.~Lo, B.~van Ginneken \emph{et~al.}, ``Extraction of airways from ct
  (exact'09),'' \emph{IEEE Transactions on Medical Imaging}, vol.~31, no.~11,
  pp. 2093--2107, 2012.

\bibitem{Unet}
O.~Ronneberger, P.~Fischer, and T.~Brox, ``U-net: Convolutional networks for
  biomedical image segmentation,'' in \emph{Medical Image Computing and
  Computer-Assisted Intervention -- MICCAI 2015}, N.~Navab, J.~Hornegger
  \emph{et~al.}, Eds.\hskip 1em plus 0.5em minus 0.4em\relax Cham: Springer
  International Publishing, 2015, pp. 234--241.

\bibitem{3dUnet}
Özgün Çiçek, A.~Abdulkadir \emph{et~al.}, ``3d u-net: Learning dense
  volumetric segmentation from sparse annotation,'' in \emph{Medical Image
  Computing and Computer-Assisted Intervention -- MICCAI 2016}, S.~Ourselin,
  L.~Joskowicz \emph{et~al.}, Eds.\hskip 1em plus 0.5em minus 0.4em\relax Cham:
  Springer International Publishing, 2016, pp. 424--432.

\bibitem{Vnet}
F.~Milletari, N.~Navab, and S.-A. Ahmadi, ``V-net: Fully convolutional neural
  networks for volumetric medical image segmentation,'' in \emph{2016 Fourth
  International Conference on 3D Vision (3DV)}, 2016, pp. 565--571.

\bibitem{CHARBONNIER201752}
\BIBentryALTinterwordspacing
J.-P. Charbonnier, E.~M. van Rikxoort \emph{et~al.}, ``Improving airway
  segmentation in computed tomography using leak detection with convolutional
  networks,'' \emph{Medical Image Analysis}, vol.~36, pp. 52--60, 2017.
  [Online]. Available:
  \url{https://www.sciencedirect.com/science/article/pii/S136184151630202X}
\BIBentrySTDinterwordspacing

\bibitem{Jin2017}
D.~Jin, Z.~Xu \emph{et~al.}, ``3d convolutional neural networks with graph
  refinement for airway segmentation using incomplete data labels,'' in
  \emph{Machine Learning in Medical Imaging}, Q.~Wang, Y.~Shi \emph{et~al.},
  Eds.\hskip 1em plus 0.5em minus 0.4em\relax Cham: Springer International
  Publishing, 2017, pp. 141--149.

\bibitem{Meng2017}
Q.~Meng, H.~R. Roth \emph{et~al.}, ``Tracking and segmentation of the airways
  in chest ct using a fully convolutional network,'' in \emph{Medical Image
  Computing and Computer-Assisted Intervention -- MICCAI 2017}, 2017, pp.
  198--207.

\bibitem{radialDistance}
C.~Wang, Y.~Hayashi \emph{et~al.}, ``Tubular structure segmentation using
  spatial fully connected network with radial distance loss for 3d medical
  images,'' in \emph{Medical Image Computing and Computer Assisted Intervention
  -- MICCAI 2019}, D.~Shen, T.~Liu \emph{et~al.}, Eds.\hskip 1em plus 0.5em
  minus 0.4em\relax Cham: Springer International Publishing, 2019, pp.
  348--356.

\bibitem{Qin2020}
Y.~Qin, Y.~Gu \emph{et~al.}, ``Airwaynet-se: A simple-yet-effective approach to
  improve airway segmentation using context scale fusion,'' in \emph{2020 IEEE
  17th International Symposium on Biomedical Imaging (ISBI)}, 2020, pp.
  809--813.

\bibitem{Zheng2021}
H.~Zheng, Y.~Qin \emph{et~al.}, ``Alleviating class-wise gradient imbalance for
  pulmonary airway segmentation,'' \emph{IEEE Transactions on Medical Imaging},
  vol.~40, pp. 2452--2462, 9 2021.

\bibitem{nnUnet}
F.~Isensee, P.~F. Jaeger \emph{et~al.}, ``nnu-net: a self-configuring method
  for deep learning-based biomedical image segmentation,'' \emph{Nature
  Methods}, vol.~18, pp. 203--211, 2 2021.

\bibitem{ViT}
\BIBentryALTinterwordspacing
A.~Dosovitskiy, L.~Beyer \emph{et~al.}, ``An image is worth 16x16 words:
  Transformers for image recognition at scale,'' in \emph{International
  Conference on Learning Representations}, 2021. [Online]. Available:
  \url{https://openreview.net/forum?id=YicbFdNTTy}
\BIBentrySTDinterwordspacing

\bibitem{Wang2021TransBTS}
W.~Wang, C.~Chen \emph{et~al.}, ``Transbts: Multimodal brain tumor segmentation
  using transformer,'' in \emph{Medical Image Computing and Computer Assisted
  Intervention – MICCAI 2021: 24th International Conference, Strasbourg,
  France, September 27–October 1, 2021, Proceedings, Part I}.\hskip 1em plus
  0.5em minus 0.4em\relax Berlin, Heidelberg: Springer-Verlag, 2021, p.
  109–119.

\bibitem{Hatamizadeh2022}
A.~Hatamizadeh, Y.~Tang \emph{et~al.}, ``Unetr: Transformers for 3d medical
  image segmentation,'' in \emph{2022 IEEE/CVF Winter Conference on
  Applications of Computer Vision (WACV)}.\hskip 1em plus 0.5em minus
  0.4em\relax IEEE, 1 2022, pp. 1748--1758.

\bibitem{Li2021}
S.~Li, X.~Sui \emph{et~al.}, ``Medical image segmentation using
  squeeze-and-expansion transformers,'' in \emph{Proceedings of the Thirtieth
  International Joint Conference on Artificial Intelligence}.\hskip 1em plus
  0.5em minus 0.4em\relax International Joint Conferences on Artificial
  Intelligence Organization, 8 2021, pp. 807--815.

\bibitem{Liu2021}
Z.~Liu, Y.~Lin \emph{et~al.}, ``Swin transformer: Hierarchical vision
  transformer using shifted windows,'' in \emph{2021 IEEE/CVF International
  Conference on Computer Vision (ICCV)}.\hskip 1em plus 0.5em minus 0.4em\relax
  IEEE, 10 2021, pp. 9992--10\,002.

\bibitem{Zhang2020}
H.~Zhang, M.~Shen \emph{et~al.}, ``Pathological airway segmentation with
  cascaded neural networks for bronchoscopic navigation,'' in \emph{2020 IEEE
  International Conference on Robotics and Automation (ICRA)}.\hskip 1em plus
  0.5em minus 0.4em\relax IEEE, 5 2020, pp. 9974--9980.

\bibitem{Zhao2020}
H.~Zhao, X.~Qiu \emph{et~al.}, ``High‐quality retinal vessel segmentation
  using generative adversarial network with a large receptive field,''
  \emph{International Journal of Imaging Systems and Technology}, vol.~30, pp.
  828--842, 9 2020.

\bibitem{Guo2020}
X.~Guo, C.~Chen \emph{et~al.}, ``Retinal vessel segmentation combined with
  generative adversarial networks and dense u-net,'' \emph{IEEE Access},
  vol.~8, pp. 194\,551--194\,560, 2020.

\bibitem{Park2020}
K.-B. Park, S.~H. Choi, and J.~Y. Lee, ``M-gan: Retinal blood vessel
  segmentation by balancing losses through stacked deep fully convolutional
  networks,'' \emph{IEEE Access}, vol.~8, pp. 146\,308--146\,322, 2020.

\bibitem{Ronneberger2015}
O.~Ronneberger, P.~Fischer, and T.~Brox, ``U-net: Convolutional networks for
  biomedical image segmentation,'' in \emph{Medical Image Computing and
  Computer-Assisted Intervention -- MICCAI 2015}, N.~Navab, J.~Hornegger
  \emph{et~al.}, Eds.\hskip 1em plus 0.5em minus 0.4em\relax Cham: Springer
  International Publishing, 2015, pp. 234--241.

\bibitem{Lee2014}
\BIBentryALTinterwordspacing
C.-Y. Lee, S.~Xie \emph{et~al.}, ``{Deeply-Supervised Nets},'' in
  \emph{Proceedings of the Eighteenth International Conference on Artificial
  Intelligence and Statistics}, ser. Proceedings of Machine Learning Research,
  G.~Lebanon and S.~V.~N. Vishwanathan, Eds., vol.~38.\hskip 1em plus 0.5em
  minus 0.4em\relax San Diego, California, USA: PMLR, 09--12 May 2015, pp.
  562--570. [Online]. Available:
  \url{https://proceedings.mlr.press/v38/lee15a.html}
\BIBentrySTDinterwordspacing

\bibitem{Juarez2018}
A.~Garcia-Uceda~Juarez, H.~A. W.~M. Tiddens, and M.~de~Bruijne, ``Automatic
  airway segmentation in chest ct using convolutional neural networks,'' in
  \emph{Image Analysis for Moving Organ, Breast, and Thoracic Images},
  D.~Stoyanov, Z.~Taylor \emph{et~al.}, Eds.\hskip 1em plus 0.5em minus
  0.4em\relax Cham: Springer International Publishing, 2018, pp. 238--250.

\bibitem{Isola2016}
P.~Isola, J.-Y. Zhu \emph{et~al.}, ``Image-to-image translation with
  conditional adversarial networks,'' in \emph{2017 IEEE Conference on Computer
  Vision and Pattern Recognition (CVPR)}, 2017, pp. 5967--5976.

\bibitem{Shit2020}
S.~Shit, J.~C. Paetzold \emph{et~al.}, ``cldice - a novel topology-preserving
  loss function for tubular structure segmentation,'' in \emph{2021 IEEE/CVF
  Conference on Computer Vision and Pattern Recognition (CVPR)}, 2021, pp.
  16\,555--16\,564.

\bibitem{nan2022}
\BIBentryALTinterwordspacing
Y.~Nan, J.~Del~Ser \emph{et~al.}, ``Fuzzy attention neural network to tackle
  discontinuity in airway segmentation,'' 2022. [Online]. Available:
  \url{https://arxiv.org/abs/2209.02048}
\BIBentrySTDinterwordspacing

\bibitem{LIDC}
S.~G. Armato, G.~McLennan \emph{et~al.}, ``The lung image database consortium
  (lidc) and image database resource initiative (idri): A completed reference
  database of lung nodules on ct scans,'' \emph{Medical Physics}, vol.~38, pp.
  915--931, 1 2011.

\bibitem{NaviAirway}
\BIBentryALTinterwordspacing
A.~Wang, T.~C.~C. Tam \emph{et~al.}, ``Naviairway: a bronchiole-sensitive deep
  learning-based airway segmentation pipeline,'' 2022. [Online]. Available:
  \url{https://arxiv.org/abs/2203.04294}
\BIBentrySTDinterwordspacing

\bibitem{Walsh2012}
S.~L. Walsh, N.~Sverzellati \emph{et~al.}, ``Chronic hypersensitivity
  pneumonitis: high resolution computed tomography patterns and pulmonary
  function indices as prognostic determinants,'' \emph{European Radiology},
  vol.~22, pp. 1672--1679, 8 2012.

\bibitem{Walsh2014}
\BIBentryALTinterwordspacing
------, ``Connective tissue disease related fibrotic lung disease: high
  resolution computed tomographic and pulmonary function indices as prognostic
  determinants,'' \emph{Thorax}, vol.~69, no.~3, pp. 216--222, 2014. [Online].
  Available: \url{https://thorax.bmj.com/content/69/3/216}
\BIBentrySTDinterwordspacing

\bibitem{Kuo2020}
W.~Kuo, A.~Perez-Rovira \emph{et~al.}, ``Airway tapering: an objective image
  biomarker for bronchiectasis,'' \emph{European Radiology}, vol.~30, pp.
  2703--2711, 5 2020.

\end{thebibliography}
%




\end{document}